# Morphological analysis of stylolites for paleostress estimation in limestones surrounding the Andra Underground Research Laboratory site

Rolland A.[(a,b)*], R. Toussaint[(a)], P. Baud[(a)], N. Conil[(b)], P. Landrein[(b)]

[(a)] Institut de Physique du Globe de Strasbourg, UMR 7516 CNRS,
Université de Strasbourg,5 rue René Descartes, 67084 Strasbourg Cedex, France; alexandra.rolland@unistra.fr ; patrick.baud@unistra.fr ; renaud.toussaint@unistra.fr
(b) Andra, Laboratoire de recherche souterrain de Meuse/Haute-Marne, RD 960, BP 9, 55290, Bure, France ; Nathalie.Conil@andra.fr; Philippe.Landrein@andra.fr

*Corresponding author. Tel.: +33-3-68-85-00-87; E-mail address: arolland@unistra.fr; Postal address: EOST, 5 rue René Descartes, 67084, Strasbourg cedex, France

**Abstract**

We develop and test a methodology to infer paleostress from the morphology of stylolites within borehole cores. This non-destructive method is based on the analysis of the stylolite trace along the outer cylindrical surface of the cores. It relies on an automatic digitization of high-resolution photographs and on the spatial Fourier spectrum analysis of the stylolite traces. We test and show, on both synthetic and natural examples, that the information from this outer cylindrical surface is equivalent to the one obtained from the destructive planar sections traditionally used. The assessment of paleostress from the stylolite morphology analysis is made using a recent theoretical model, which links the morphological properties to the physical processes acting during stylolite evolution. This model shows that two scaling regimes are to be expected for the stylolite height power spectrum, separated by a cross-over length that depends on the magnitude of the paleostress during formation. We develop a non linear fit method to automatically extract the cross-over lengths from the digitized stylolite profiles. Results on cores from boreholes drilled in the surroundings of the Andra Underground Research Laboratory located at Bure, France, show that different groups of sedimentary stylolites can be distinguished, and correspond to different estimated vertical paleostress values. For the Oxfordian formation, one group of stylolites indicate a paleostress of around 10 MPa, while another group yields 15 MPa. For the Dogger formation, two stylolites indicate a paleostress of around 10 MPa, while others appear to have stopped growing at paleostresses between 30 and 22 MPa, starting at an erosion phase that initiated in the late Cretaceous and continues today. This method has a high potential for further applications on reservoirs or other geological contexts where stylolites are present.

## 1. Introduction

Pressure-solution is a complex process that results in strain localization (localized dissolution) under particular stress conditions, and is at the origin of stylolite formation in several types of sedimentary rocks (e.g., carbonates ([1]-[5]), sandstones [6], and shales [7]). Stylolites have an undulated shape and are filled with organic matter, oxides, or clay particles. The latter has a significant role in the kinetics of the nucleation process ([8], [9]). According to Fabricius and Borre [10], it is the burial stress that controls the pressure-solution process, while the temperature controls recrystallisation and cementation. Stylolites can be divided in several families, according to their orientation. Bedding-parallel stylolites are called sedimentary stylolites and differ from tectonic stylolites that form, in most cases, at a high angle to the bedding (i.e., sub-vertical). Sedimentary stylolites form due to lithostatic pressure, while tectonic stylolites form due to major compressive stresses related to the tectonic stress field (the Alps, the Pyrenees [11]), i.e., when the largest principal stress becomes horizontal rather than vertical. Because of the different morphologies and orientations associated with the stress conditions in which they developed, previous studies ([12], [13], [14]) suggested that the stylolite morphology contains a signature of the stress field that can be extracted from its roughness. A few studies have tried to reproduce stylolites by simulating the pressure-solution process in the laboratory. However, such studies are very difficult due to the slow kinetics of the process [7]. While some studies were based on loading aggregates in the presence of either saturated or non-saturated fluids ([9], [15], [16], [17]), others used an indenter to load crystals or surfaces, in the presence of fluid, whilst monitoring their evolution with time ([18], [19], [20], [21]). However, whilst one study produced microstylolites at the stressed contacts between grains [17], most experiments yielded other microstructural features, such as grooves ([22], [23], [24]). To our knowledge, stylolite growth has not yet been observed unambiguously in the laboratory. The study of the occurrence of stylolites, and their potential use as geological markers, therefore largely relies on the success of theoretical modelling. Numerical and analytical models have therefore been developed to study the growth of stylolites in presence of clay [1], or to reproduce the roughening from a preferential existing surface ([21], [25], [26]). Renard et al. [12] and Schmittbuhl et al. [13] observed from morphological analyses of the stylolite roughness, two regimes depending on different range of scales separated by a characteristic length. Schmittbuhl et al. [13] proposed a short version of an analytical model that predicts the growth of a stylolite from a fluid-solid interface to describe these observations. The details of the model were published recently by Rolland et al. [27]. It takes into account some considerations about mechanical and chemico-mechanical equilibrium to express the dissolution of the rock mass at a fluid-solid interface which mimic in a simple way the initial stage of the stylolite growth. This model gives a relation between the applied stress during the stylolite development and a characteristic length associated with the stylolite morphology. This relation was supported by numerical models studying the effect of disorder on the evolution of stylolite morphology ([27], [28]), and by some pilot studies performed on natural stylolites sampled at various depths at the same site [14]. This approach therefore provides a tool to infer the stress history in various geological environments where stylolites are present.

Such an analysis hinges of course on a detailed description of the stylolite morphology, which to date has only been conducted in a few pilot studies ([13], [29]), on either 1D profiles or 2D surfaces. These analyses used digitized stylolite profiles or elevations, and were carried out on carbonate formations within newly-opened quarries or outcrops in the Cirque de Navacelles (Massif Central), Burgundy and Jura, Chartreuse, and Vercors mountains of France. In these studies, a characteristic length called the cross-over length (typically around the millimetre scale) was extracted by analysing the stylolitic profiles or surface height variations over different scales. The cross-over length separates the two scaling regimes predicted by the analytical model [27] presented before. On the small-scale, surface tension is the dominating process; while, on the large-scale, the roughness is driven by elastic interactions. In most of the geophysical/geological applied problems (such as reservoir/aquifer management, and nuclear waste repository management), the stress history of the sites is a fundamental issue. The systematic determination of the paleostress using stylolite morphology can therefore become an important tool in various applied contexts. This type of analysis demands core samples extracted *in-situ* from non-destructive boreholes (i.e., those where full core samples are retrieved) drilled over a representative depth interval within the target reservoir or aquifer. Obviously, these borehole samples need to contain stylolites. However, the use of cores from boreholes induces some geometrical limits linked, in particular, to their finite size and cylindrical shape. To our knowledge, only one example of such an analysis has been presented so far, on a core taken near from the Andra Underground Research Laboratory (URL) at Bure from the Dogger formation [27].

The objective of this study is to test systematically the applicability of this method using a large number of stylolites within cores from borehole samples, and to develop a rigorous methodology to deduce the stress history of a reservoir using stylolite morphology. A previous study [30] showed the extensive presence of stylolites at various depths around the Andra URL at Bure. Therefore, this context is an excellent candidate to apply our methodology for the assessment of the paleostresses recorded by the stylolite morphology.

## 2. Geological context of the Bure URL

The Andra URL is located at Bure in the eastern part of the Paris Basin in France. Since 2001, it has been designed to study the feasibility of a nuclear waste repository in the Callovo-Oxfordian

(COX) claystones. The target horizon is surrounded by limestones from the Oxfordian and Dogger ages. Based on stress measurements and estimations around the Andra URL, Gunzburger and Cornet [31] suggested that the observed differential stress in the COX claystone formation (Fig. 1) is due to pressure-solution acting in the surrounding Oxfordian and Dogger limestone formations. Indeed, this differential stress implies the existence of an active process occurring in the surrounding formations. The presence of numerous stylolites within these limestone formations supports the fact that this process is actually pressure-solution.

**Figure 1:**

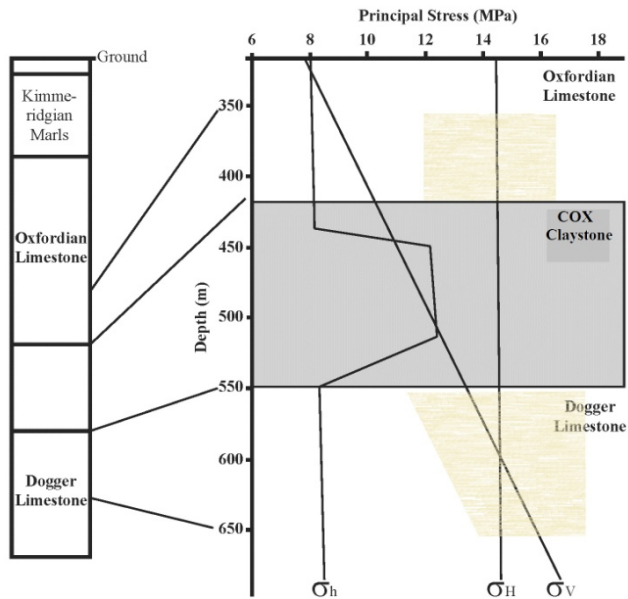

Figure 1: Estimated stress profiles at the URL based on in-situ stress measurements (modified from Gunzburger and Cornet [31]). $\sigma_V$ is the major principal vertical stress and $\sigma_H$ and $\sigma_h$ are the maximum and minimum principal horizontal stress, respectively. Gunzburger and Cornet [31] suggests that the observed differential stress in the Callovo-Oxfordian claystone formation is due to slow processes, such as pressure-solution, acting in the surrounding limestone formations. The light four-sided shapes represent the error bars on the $\sigma_H$ estimation.

Boreholes were drilled adjacent to the Andra URL to study three horizons. However, from the 44 drilled boreholes, only 10 were partially non-destructive: 6 within the COX formation, 2 within the Oxfordian formation (EST204 and EST205), and 2 within the Dogger formation (EST433 and EST210). All of these 10 boreholes were drilled vertically (i.e., perpendicular to the surface and the bedding). Cores were recovered from the surface to depths of 508 and 510 m from boreholes EST204 and EST205, respectively. Cores between the depth of 526 to 770 m were recovered from borehole EST433, with partial core recovery to a depth of 2001 m. Since the aim of this study is to sample stylolites over the largest depth interval possible, we selected the boreholes EST205 and EST433 as they presented the highest potential in terms of available cores.

The sampling of the stylolites within the cores is not regular. As we aim to describe the morphology for each selected stylolite, we need, in the available cores, isolated stylolites with no interactions with other stylolites or big heterogeneities to avoid noise in the data. Different zones were encountered within the borehole core samples: some presented regularly spaced stylolites (Fig. 2a) that we used in this study, others presented damaged zones and multiple stylolites (Fig. 2b), and some zones present no stylolites. Additionally, stylolites that contained a fracture in their seam and slickolites (stylolites with tilted teeth) were not used in our study. Anastomosing stylolites were frequently observed in the studied cores, but were not suitable for the morphological analysis as they split in different branches and the theoretical model applied in this study only considers the growth of an isolated stylolite (see section 3). The above-mentioned stylolites were excluded from our dataset. A certain number of petrophysical measurements were recently performed on the cores studied here [32]. More specifically, we used uniaxial tests to determine the Young's modulus, *E*, that will be used to assess the paleostress associated to each stylolite.

**Figure 2:**

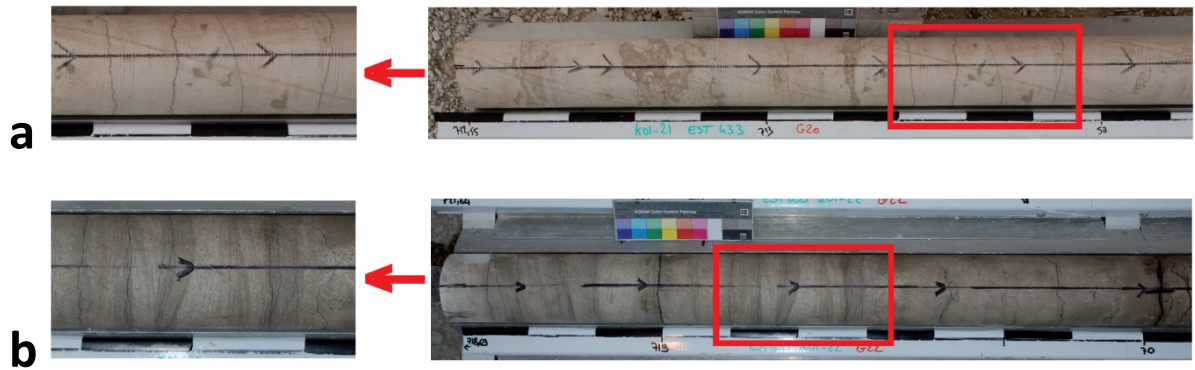

Figure 2: Photographs of the studied material. a) Core from the EST433 borehole (712.55 to 713.60 m) with several distinct stylolites. The left image is a zoom of the red box. b) Core from the EST433 borehole (718.69 to 719.75 m) with a zone of multiple stylolites and damaged parts. The left image is a zoom of the red box.

Our initial goal was to analyse already-opened stylolites and stylolites opened in the laboratory (Fig. 3). Roughness can be quantified precisely for these stylolites, at resolution down to a few tenths of micron using laser profilometry ([11], [12], [13], [21], [33]). Obviously, this can only be done if enough open stylolites are available and/or if one can open enough stylolites. In the selected boreholes, as in the surrounding ones, open stylolites were scarce. Moreover, they often appear to have been weathered and hence affected by processes that could alter their morphology (Fig. 3a). Further, although it is sometimes possible to open a stylolite in the laboratory, significant fracturing usually occurs for stylolites that contain only a thin clay layer. Laser profilometry would thus scan a combination of stylolite and fracture. An example of this situation is given in Fig. 3b where a 4 cm x 2 cm sample broke partially along the stylolite and a fracture-like structure emerged. Therefore, to obtain the most-reliable results, we adopted a methodology that uses closed stylolites only (i.e., on 1D stylolitic profiles). To achieve our goal, a dedicated procedure was developed and will be outlined in section 4.

**Figure 3:**

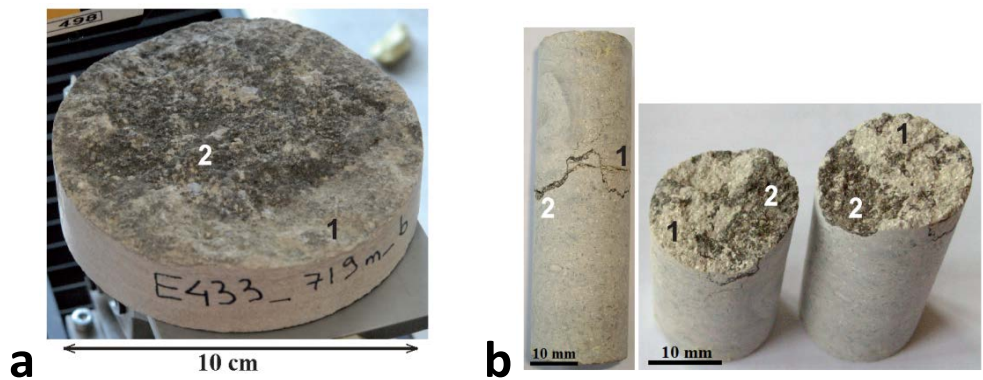

Figure 3: The issue of open and opened stylolites. a) Open stylolite taken from the borehole EST433 at 719 m deep with weathered surface. b) Oxfordian limestone sample (4 cm*2 cm) opened in the laboratory. We observe a combination of fracture-like structure (#1, light colour corresponding to the surrounding rock) and of stylolite-like structure (#2, dark colour corresponding to the insoluble matters in the stylolitic seam). We observe similar features on the sample coming from the borehole EST433.

3. **Theoretical background**

To better understand how stylolite growth can be described, we will briefly review the model of Rolland et al. [27]. Using a simplified initial geometry (elongated fluid pocket enclosed between two contactless surfaces of infinite extent, Fig. 4), the manner in which the dissolution speed at the fluid-solid interface is affected by heterogeneities (by taking into account mechanical equilibrium and chemo-mechanical coupling) can be calculated. The surface is forming in a far-field stress tensor $\overline{\overline{\sigma^0}}$ where the horizontal principal components are isotropic ($\overline{\overline{\sigma^0_{xx}}} = \overline{\overline{\sigma^0_{yy}}}$) and smaller than the vertical principal component $\overline{\overline{\sigma^0_{zz}}}$. First, they show that the mechanical equilibrium at the solid-fluid interface is $\sigma.\hat{n} = -p\hat{n}$ where $p = -\sigma^0_{zz}$ is the fluid pressure. A local stress perturbation $\overline{\overline{\sigma^1}}$, induced by the irregularities of the surface, is combined to the far-field stress $\overline{\overline{\sigma^0}}$ and gives the stress field at the interface. They deduced the force perturbation due to the local stress perturbation $\overline{\overline{\sigma^1}}$ with $\delta f(x) = \sigma^1(x).\hat{n} = \sigma^0_s(\partial_x z)\hat{x}$ where $\sigma_s = \sigma_{zz} - \sigma_{xx}$ is the differential stress. The chemo-mechanical coupling is expressed by the calculation of the dissolution speed normal to the solid-fluid interface as $v = m\Delta\mu$ where $m$ is the mobility of the dissolving phases depending on the dissolution rate $k$ and molar volume $\Omega$ of the phases at a given temperature, and $\Delta\mu$ is the chemical potential depending on the Helmoltz free energy, the change in normal stress, and on the curvature $\kappa = \partial_{xx} z$ of the interface. As for mechanical equilibrium, the velocity is affected by the perturbation of the surface giving a dissolution speed $v = v^0 + v^1$. Expressing the chemical potential as a function of the elastic free energy gives $u_e = \frac{[(1+v)\sigma_{ij}\sigma_{ij} - v\sigma_{kk}\sigma_{ll}]}{4E}$, where $E$ is the Young's modulus and $v$ the Poisson's ratio. The dissolution speed is expressed as $v = -\partial_t z = m\Omega(\Delta u_e + \gamma\kappa)$ where $\gamma$ is the surface tension. The calculation of the strain $\epsilon_{ij}$ and of the stress $\sigma_{ij} = \frac{\left[\epsilon_{ij} + \left(v\epsilon_{kk}\delta_{ij}/1-2v\right)\right]E}{1+v}$ using the Green function method (see details in [27]) permits the calculation of the stress perturbation $\sigma^1$ induced by the force perturbation $\delta f(x)$ and the elastic energy perturbation $u_e^1$. Combining all these equations yields an expression for the dissolution speed:

$$\partial_t z = \partial_{xx} z - \frac{l}{L_c}\int \frac{\partial_y z}{x-y} dy + \eta \qquad (1)$$

where $\eta$ is a quenched noise due to the rock heterogeneities and $L_c = \frac{\gamma E}{\beta P \sigma_s}$ is a characteristic length where $\beta = \frac{[2v(1-2v)]}{\pi}$ is a dimensionless constant and $P$ is the mean pressure. This equation describes the dominating process at both the small- and large-scale (i.e., at lengths inferior or superior to the characteristic length $L_c$). At the small-scale, i.e., $l \ll L_c$, surface tension is the dominating process and the model is reduced to $\partial_t z = \partial_{xx} z + \eta$.

**Figure 4:**

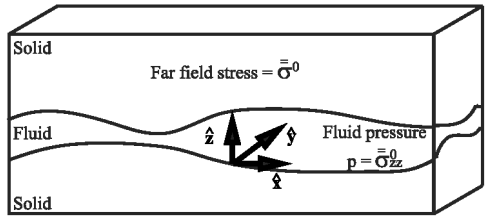

Figure 4: Geometry of the solid-fluid interface modified from Rolland et al. [32]. An elongated pore fill with fluid is embedded in the solid matrix on which a global stress field $\sigma^0$ is applied.

At the large-scale, i.e., $l \gg L_c$, elastic interactions dominate and the model is reduced to $\partial_t z = -\dfrac{l}{L_c} \int \dfrac{\partial_y z}{x-y} dy + \eta$. Thus, this model highlights the occurrence of two regimes driven by two different processes. These two processes lead to two different scaling laws driven by $L^{(2H+1)/2}$ where $H$ is the Hurst exponent which is close to 1 for the small-scale ([13], [34]) and to 0.5 for the large-scale [35]. Both regimes are separated by a characteristic length separating two scaling domains. This characteristic length appears in the morphology analysis presented in the next section and is called cross-over length. The cross-over length contains a signature of the stress field during the formation of the stylolite.

## 4. Morphological analysis

As we have a large number of available stylolites in the cores, we developed a semi-automatic method to analyse our samples. The idea is to have a systematic procedure that can be applied when a lot of cores are available.

### 4.1. *Procedure to digitize stylolites*

The morphology analysis requires the extraction of a profile from the stylolite which has to be a single-valued function. The recipe for the digitization (the procedure is the same for slices or external profile) are listed below and illustrated in Fig. 5:

Step i: We took the core containing the stylolite and we placed it on a circular platen that can rotate around a precisely fixed axis. We used a digital single-lens reflex camera (©NIKON D700) equipped with a 105 mm micro lens that allows us to have a high depth of field, and thus to have the curved surface in focus. We zoomed in to have the best resolution, which was 30 µm per pixel for our photographs (Fig. 5a). The sample was illuminated by two powerful spotlights on the front side to have the same illumination over the entire surface.

Step ii: To merge the pictures we used a standard graphics editing program (©Adobe Photoshop). We used grey level pictures (8-bits or 256 values) as we wanted to isolate the stylolite seam which appeared as black pixels. However, we had to clean the vicinity of the stylolite by removing some dark patches (Fig. 5b). Indeed, impurities or clay particles were often present in the rock and can be confused with the stylolite in the next step. Thus we surrounded the vicinity of the stylolite with white pixels with the stylolite appearing in black.

Step iii: We applied a threshold to isolate the black pixels constituting the stylolite seam from the surrounding rock (Fig. 5c). This threshold needed to be slightly modified for different samples according to the rock composition and to the illumination. It was done manually. Indeed, all the cores did not have the same colour, the stylolites varied in thickness depending on the quantity of insoluble material, and heterogeneities were not always present in similar quantities. In our case, the threshold was set between 60 and 110 over 256 values. The sensitivity of the results to this threshold was thoroughly tested (see Appendix 1).

Step iv: However, as mentioned in step ii, impurities can still be present and to remove them we applied another threshold on the size of the black pixels cluster (Fig. 5d). The smallest clusters, associated to the impurities, where then ignored.

Step v: As we need a single-valued function for the analysis, we interpolated the discontinued components (Fig. 5d, red segments). We chose a linear interpolation between these clusters. We demonstrate in Appendix 2 that this interpolation has no incidence on the final results.

Step vi: We extracted functions associated with the stylolite roughness (Fig. 5e). Three functions could be extracted: the top and bottom edges of the clusters, and the average line of the clusters. Our procedure to analyse the average line is presented in the next section.

**Figure 5:**

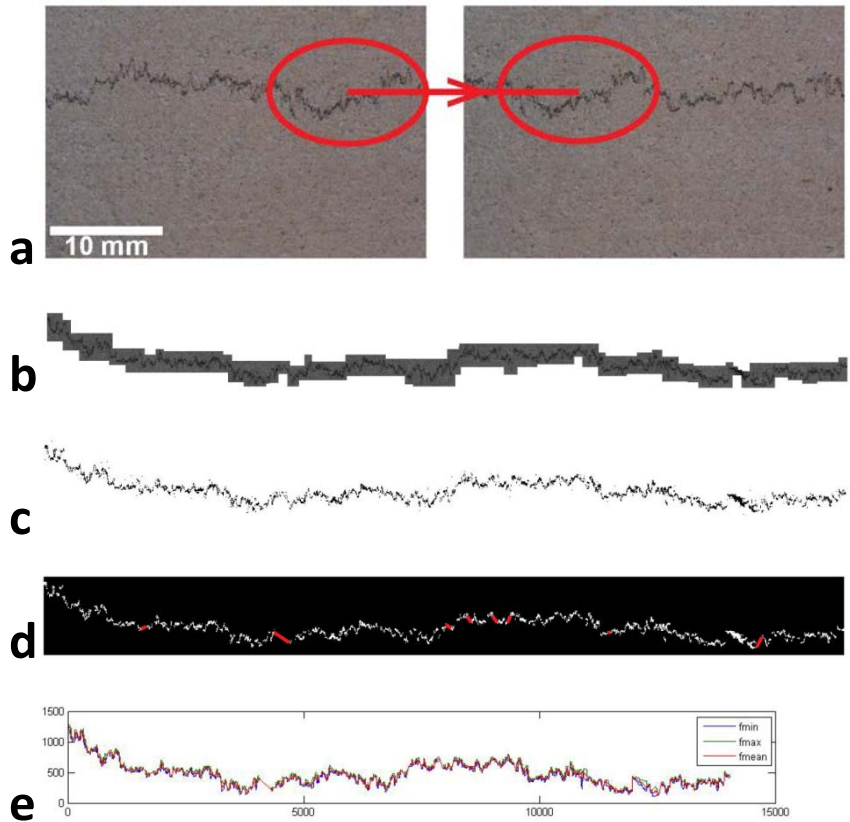

Figure 5: Main steps to digitize stylolites. a) Step i - Take pictures with high resolution. b) Step ii - Merge and clean pictures. c) Step iii - Isolate the stylolite – First threshold on the grey-level. d) Step iv and v - Isolate the stylolite – Second threshold on the cluster size and interpolate the discontinue parts. e) Step vi - Extract the functions associated with the stylolite. Axes are in pixels.

*4.2. Procedure to analyse the stylolite morphology*

Stylolites show a self-affine geometry giving them the property to be invariant by affine transformation. For the horizontal variations $\Delta x$ and $\Delta y$ and the vertical variation $\Delta z$, the self-affinity can be defined as ([36], [37]): $\Delta x \rightarrow \lambda \Delta x$, $\Delta y \rightarrow \lambda \Delta y$ and $\Delta z \rightarrow \lambda^H \Delta z$. $H$ is the Hurst exponent or roughness exponent which varies roughly between 0 and 1. Previous studies have analysed sedimentary stylolitic profile variations over different scales ([13], [14]). The results show two distinct scaling regimes, corresponding to different power laws. The exponent of these power laws is a function of the Hurst exponent $H$. Both regimes are separated by a characteristic length $L_c$, called cross-over length, typically within the millimetre scale. The small-scale regime shows a Hurst exponent around 1 and the large-scale regime has a Hurst exponent around 0.5.

To analyse the profiles, several signal processing methods exist. The main ones used to analyse stylolite or fracture morphologies are:

- the wavelet power spectrum (WPS) method ([12], [37], [38]) consisting of reconstructing the signal as a sum of different wavelets. It starts with a mother function which can be translated or dilated to find the corresponding form in the signal.
- the Fourier power spectrum (FPS) method ([12], [38]) consisting of analysing the wavelengths in the signal and reconstructing it as a sum of cosines and sines.
- four other methods – the root mean square (RMS), the maximum-minimum height difference (MM), the correlation function (COR), the RMS correlation function (RMS-COR) – allow the analysis of stylolitic signals by analysing the height variations of the signal (for details refer to Candela et al. [38]).

Candela et al. [38] ran tests on synthetic anisotropic self-affine surfaces to assess the robustness of each method. They found that the RMS-COR, the FPS, and the WPS techniques are the most reliable. We used the FPS technique to perform the spectral analysis of our stylolitic profiles. The spectrum was obtained by computing the squared Fourier transform modulus $P[k]$ of the profile as a function of the wave-number $k$ ($k = 2\pi/L$ where $L$ is the wave-length).

Considering the self-affinity geometry of the stylolite, the FPS can be expressed as a function of the Hurst exponent, and by using the wave-length, yielding: $P[L] \approx L^{2H+1}$. If we plot the FPS as a function of $L$ on a log-log graph, the spectrum is noisy (Fig. 6) with a lot of data at the small-scales and less and less data at the large-scales. To improve the analysis of the data, we resampled the data using logarithmic binning, giving a dot every 1.5 decade. This has the advantage to give a scale range of equal size in the logarithmic space.

**Figure 6:**

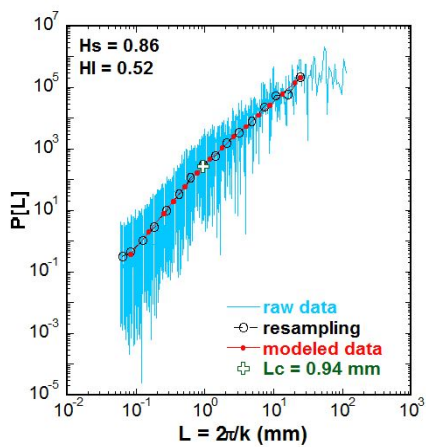

Figure 6: Spectrum obtained after the analysis of a stylolitic profile. The sample comes from the borehole EST433 at a depth of 720 m, in the Dogger formation. The Fourier power spectrum is represented as a function of the wave-length. The noisy continuous line represents the raw spectrum obtained after the analysis. The discontinuous line with open circles represents the logarithmic binning, the data being resampled every each 1.5 tenth of order. The continuous line with filled circles corresponds to the modeled data. The cross is the estimated cross-over length calculated by the code. The estimated Hurst exponents for small and large scale (*Hs* and *Hl* respectively) are in the up-left corner.

To find the cross-over length we used an automatic least-square non-linear fit method as outlined in Ebner et al. [14]. It consists of fitting the resampled data in bilogarithmic space using the least-squares method. In this space, the model (two power laws with a cross-over to be determined) corresponds to a linear function over two parts with a cross-over function changing the scaling law from the small- to the large-scale. We look for a Hurst exponent around $0.5 \pm 0.2$ and $1 \pm 0.2$ for the large-scale and small-scale, respectively. Usually, to have a better measurement, more than an order of magnitude around the cross-over length is necessary. For a cross-over length around 1 millimetre, we need at least a 10 centimetre-sized sample and a minimum resolution of the pixel size. Then, the estimation of the cross-over length is iterative.

The uncertainty in the determination of the cross-over length, noted *Lc* in the following, was assessed to be 23.34 % by the use of synthetic stylolitic 1D profiles. The synthetic profiles were obtained using the same procedure as in [39] and [40]. The uncertainties for small- and large-scale Hurst exponents were also assessed and are equal to 2.45% and 9.34%, respectively [41]. The repeatability of the method from the digitization to the spectral analysis was rigorously tested (see Appendix 3).

*4.3. Data selection*

We can extract two types of profile from the borehole core samples: planar profiles taken from a cut in the longitudinal direction of the core, and external profiles taken from the outer cylindrical surface of the core. Longitudinal slices required a long and destructive preparation process. Since a non-destructive method is needed in a variety of applications, we examine the possible distortion associated with the curvature of the cylindrical profiles by comparing the analysis of planar and cylindrical profiles. We analysed the external stylolitic profile and four planar profiles extracted from slices cut a few centimetres apart (Fig. 7a) of the same core coming from the Dogger horizon. The profiles were digitized using the procedure described above (Fig.7b). We then plotted the FPS as a function of the length *L* in a bilogarithmic scale (Fig. 7c). Considering the error bars, the results show that the cross-over lengths are very close (except for profile 1). This small discrepancy can be explained by the fact that planar profiles are shorter than external profiles, and therefore the large-scales were less well-defined by the data. The low resolution at the large-scales made the planar profiles analysis slightly more challenging. This issue was circumvented by making an average over several close planar profiles. Another difference lies in the periodic nature of the external profiles, contrarily to the planar ones. This periodicity certainly affects the determined Fourier amplitudes associated to the largest modes (around 1/4 of the perimeter or more), which should be excluded from the FPS analysis to provide a similar measure to the classical planar profiles. These possible differences were studied by performing the same analysis on synthetic data. The results are presented in Appendix 4. Both analyses showed that using the external profile or planar profiles did not seem to induce significant artefacts in the obtained cross-over lengths. Therefore, external profiles were used in the following analyses. Moreover, the analysis of sedimentary stylolites for different azimuths (Appendix 5) reveals that there is no anisotropy of the cross-over length.

**Figure 7:**

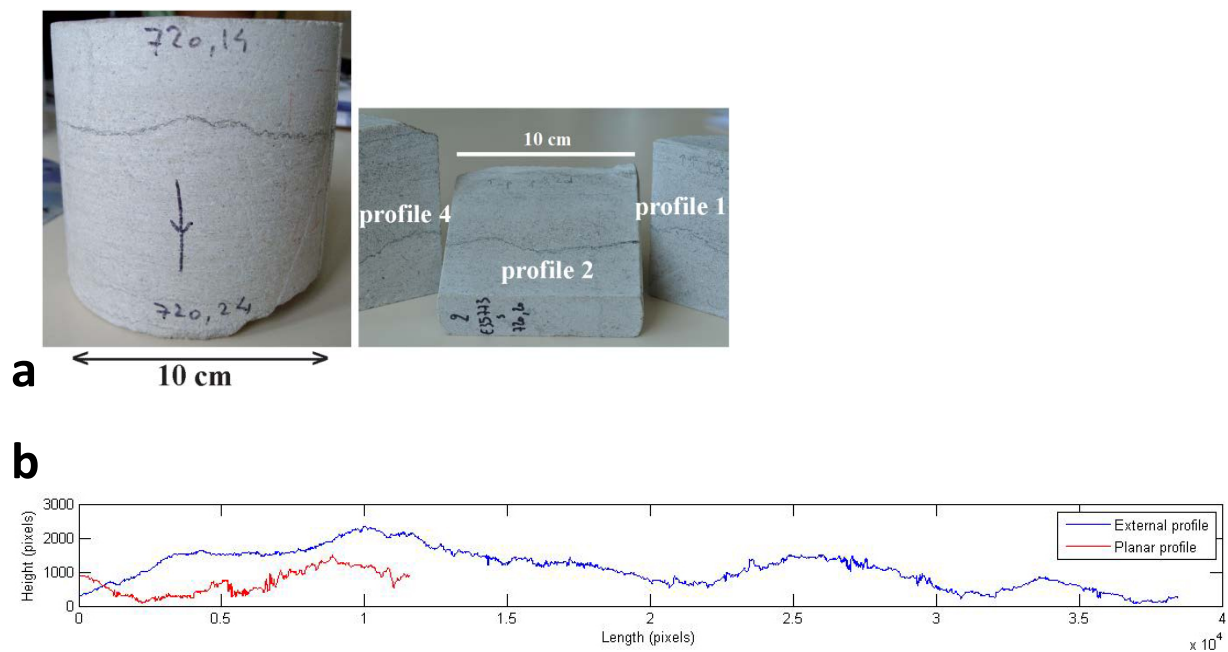

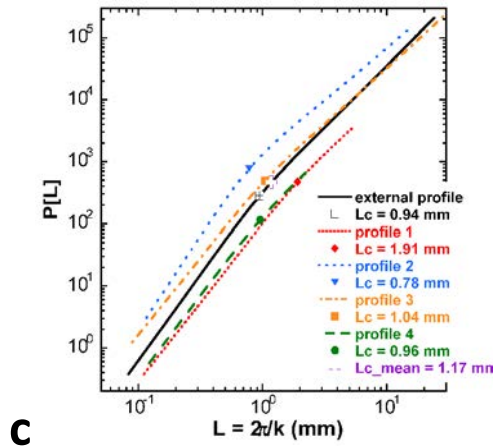

**c**

Figure 7: Comparison of the planar and external profiles spectrums. a) The analysed stylolite comes from a 10 cm diameter core taken from the borehole EST433 at 720 m deep. Four planar profiles were obtained by cutting the core in 3 slices while the external profile was extracted from the outer part of the core. b) Digitized external profile and planar profile 2. The digitized profiles do not have the same length. The resolution (pixel size) is 35 μm. c) Result of the FPS analysis. The Fourier power spectrum is represented as a function of the wave-length. The spectrum and cross-over lengths obtained are very similar.

## 5. Results and discussion
### 5.1. Sedimentary stylolites

We sampled stylolites from the Oxfordian and Dogger formation at depths ranging from 150 m to 320 m and from 650 to 750 m, respectively. 22 stylolites were selected from the Oxfordian formation, and 21 from the Dogger formation. Some of the stylolites displayed one scaling law, i.e., they could be described by a single Hurst exponent, lying typically between 0.6-0.8 (8 for Oxfordian limestones and 10 for Dogger limestones). Fig. 8a shows one example of a stylolite that could be characterised by one scaling law. In many cases, the implied stylolites had a thick seam of insoluble materials that protruded from the surface of the core, likely due to soft material rearrangement along external profiles during the cutting. This influenced the small-scale by obscuring the fine details of the morphology. Alternatively, changes in the stress field could also have erased the typical morphology and produced stylolites exhibiting one scaling law.

**Figure 8:**

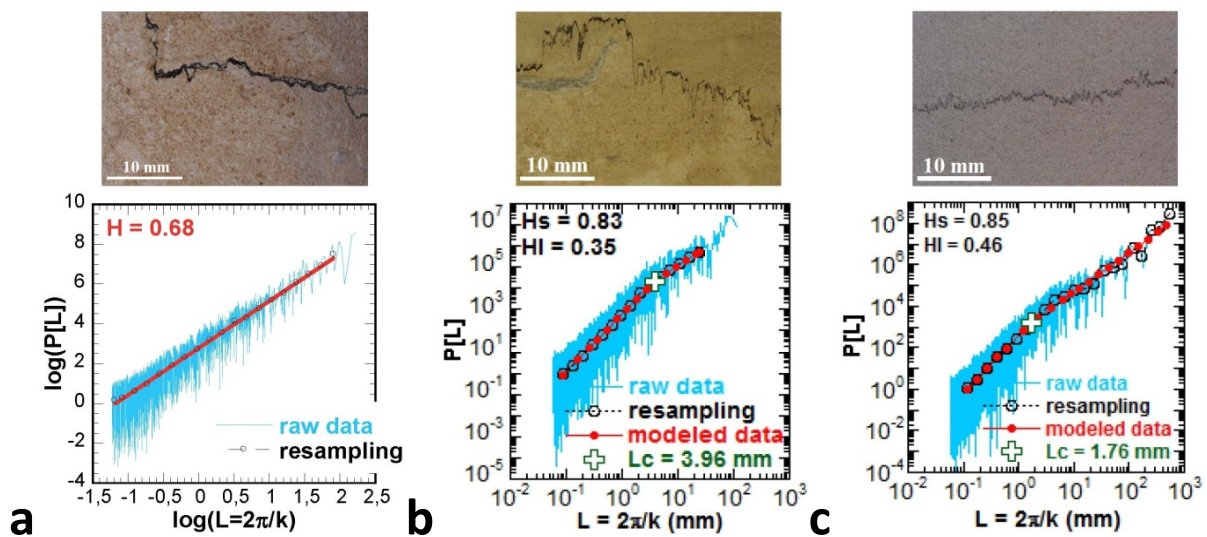

Figure 8: Representative examples of spectrums resulting from the analyses of stylolites in Oxfordian limestones. The Fourier power spectrum is represented as a function of the wave-length. a) Sample from core EST43790 at 231.40 m shows a behaviour with one scaling law described by one Hurst exponent H in red. b) Sample from core EST06683 at 159.23 m shows the two regimes behaviour. However, this sample was disregarded due to the poor quality of the fit at the large scales. c) Sample from core EST43792 at 253.92 m shows the two-regimes behaviour and is typical of the samples used to infer the paleostresses.

All the other stylolites show two scaling law regimes (16 and 11 for Oxfordian (Fig. 8b, 8c) and Dogger, respectively). Among the stylolites showing this behaviour, we observed some ill-defined spectrums, i.e., the fitted data for the small- and large-scale (Fig. 8b) did not include enough data (less than one logarithmic order of magnitude). The high resolution photographs of these stylolites did not show any obvious differences with the ones showing the two regimes. Moreover, some spectrums showed the possibility of a higher cross-over length in the large-scale but, in all cases, the right-part of the spectrums did not have enough data (3 or 4 points) to maintain the existence of a cross-over length. This was due to the geometrical limitation of our work on cores that have an inherently finite size. A recent study [21] suggests that a cross-over length could exist at a higher scale than our investigation limits imposed by the core size. This could explain why a certain number of spectra are ill-defined in the large scales. The possible existence of this cross-over length is anyway not described by the analytical model of Rolland et al. [27] and thus the morphology analyses of these stylolites cannot be carried on.Therefore, we chose to exclude the cross-over lengths when the spectrums were ill-defined. Our results (Fig. 9a) provided cross-over lengths ranging from 0.45 mm to 4 mm for the Oxfordian stylolites, and from 0.8 mm to 6.2 mm for the Dogger stylolites. The normal distributions associated to the results are plotted in Fig. 9b and Fig. 9c for Oxfordian and Dogger, respectively. We observed no systematic variations of the cross-over length with depth. This means that the stylolites do not show the same growth history. Indeed, some may have started to grow at different ages; others may have stopped growing after a particular event such as the closing of the porosity due to recristallization, and thus the cessation of the fluid flow. The history of the studied stylolites in these long series is thus more variable than in simpler contexts, such as in the Cirque de Navacelles [14] where a linear trend between the cross-over length and the depth was observed. However, we can distinguish several groups of stylolites. For the Oxfordian stylolites, there are two groups: one with cross-over lengths around 1.2 mm, and one around 3 mm. The Dogger stylolites could be divided into three distinct groups, characterised by the cross-over length: one around 1 mm, one around 2.3 mm, and one around 5 mm.

**Figure 9:**

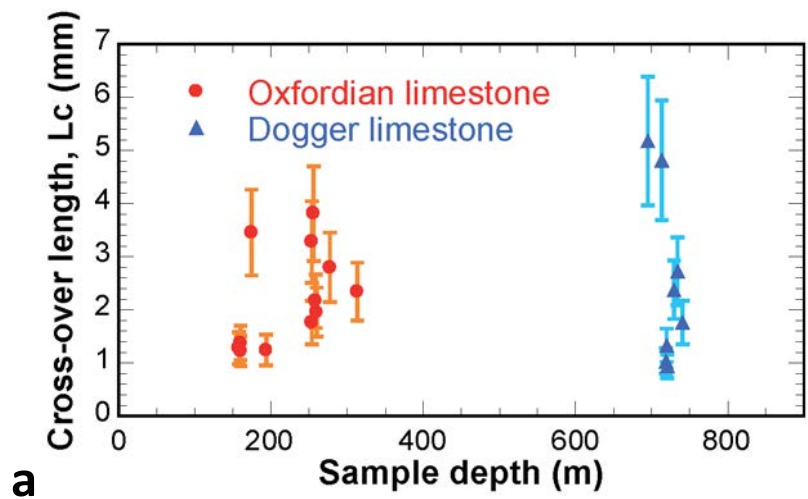

a

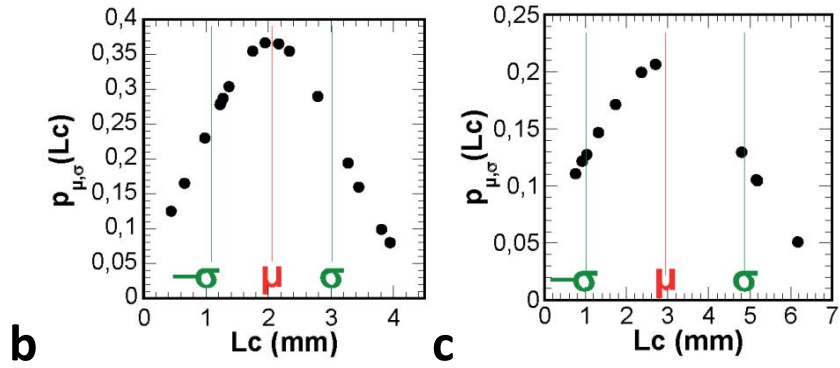

Figure 9: Summary of the estimated cross-over lengths. a) Cross-over length as a function of the sample depth for the stylolites from the Oxfordian and Dogger formations. The error bars represent 23.34% of the value. b) Normal distribution of the resulting cross-over lengths for the Oxfordian limestones. c) Normal distribution of the resulting cross-over lengths for the Dogger limestones. Some of the extreme values were excluded because of ill-defined spectrums.

We calculated the corresponding paleostress using Equation (1) where $L_c = \frac{\gamma E}{\beta P \sigma_s}$. This equation can be simplified based on assumptions concerning the geological context. We applied the same simplification as in Rolland et al. [27] to the basin evolution which is assumed to be at its early stage when the first stylolites initiate. As we are in a young depositional context, the major principal stress is assumed to be vertical, $\sigma_{zz}$, the horizontal principal stresses are assumed to be isotropic, $\sigma_{xx} = \sigma_{yy}$, and the strain is assumed to be uniaxial and vertical, $\sigma_{xx} = \sigma_{yy} = \frac{v}{1-v}\sigma_{zz}$. Using these assumptions, we have:

$$P = (2\sigma_{xx} + \sigma_{zz})/3 \qquad (2)$$

and,

$$\sigma_s = \sigma_{zz} - \sigma_{xx} \qquad (3)$$

And therefore, using Equation (1-3),

$$\sigma_{zz}^2 = \frac{\gamma E}{\alpha \beta L_c} \qquad (4)$$

where $\alpha$ is a dimensionless geometrical factor depending on the Poisson's ratio $v$ ($\alpha = \frac{1}{3}\frac{(1+v)(1-2v)}{(1-v)(1-v)}$).

Using Equation (4), we assessed the paleostress $\sigma_{zz}$ for sedimentary stylolites from the estimated cross-over length $L_c$. The results are presented on Fig. 10. For the calculation, we used $\gamma = 0.27$ J.m$^{-2}$ which is the surface tension for a calcite-water interface [12]. X-ray diffraction measurements show a composition comprising of at least 97 % calcite. Hence, we used the Poisson's ratio of calcite $v = 0.32$ [42]. Considering the error induced by the cross-over length estimation, the calculated error on the paleostress was 11.67 %. Regarding the Young's modulus, as we can only measure the present value, we considered two end members for the Young's modulus assuming two different scenarios for the stylolite growth. In the first scenario, the stylolite evolution stopped when the conditions for pressure-solution were not fulfilled anymore (closure of the pores and thus decrease of the dissolution process) or when there was a significant change in the stress field. The Young's modulus is considered in this case to be different from the present value, i.e., corresponds to the value at the end of the stylolitization process for which we have no clue about the age at this stage. Thus, we defined the lower bound for the Young's modulus using the lowest published values for limestone, E = 15 GPa [43]. In the second scenario, the stylolite is still evolving so the Young's modulus is considered to be the same as the present value and defines the upper bound. We used the values measured in the laboratory [32] ranging from 23 to 36 GPa for the Oxfordian formation, and from 40 to 80 GPa for the Dogger formation. Young's moduli were measured using the uniaxial and triaxial presses of the

*Laboratoire de Géophysique Expérimentale* of *EOST Strasbourg*. Details about the experimental set-up can be found in Heap et al. [44] and Baud et al. [45], respectively. Loading-unloading cycles were performed on each sample and *E*, the so called tangent modulus, was measured during the unloading [46]. For samples at depth different from horizons where measurements were done, we extrapolated the Young's modulus by taking the Young's modulus of the nearest horizon. These two end-members scenarios resulted in the estimates of the paleostresses shown on the Fig. 10a and 10b. As a reference, we added two curves representing (i) the present lithostatic stress field ($\sigma_{zz} = \rho g z$ where $\rho$ = 2700 kg.m$^{-3}$ is the density, $g$ = 9.81 m.s$^{-2}$, and *z* is the depth) and, (ii) the stress field at the maximum overburden that both limestone formations have seen. It was shown in a recent study [47] that the maximum erosion thickness that occurred during the late Cretaceous period is around 320 m. Regarding the samples from the borehole EST205, a thickness of 120 m has to be added to the overburden because of a localised erosion phase of the Barrois limestones [47].

**Figure 10:**

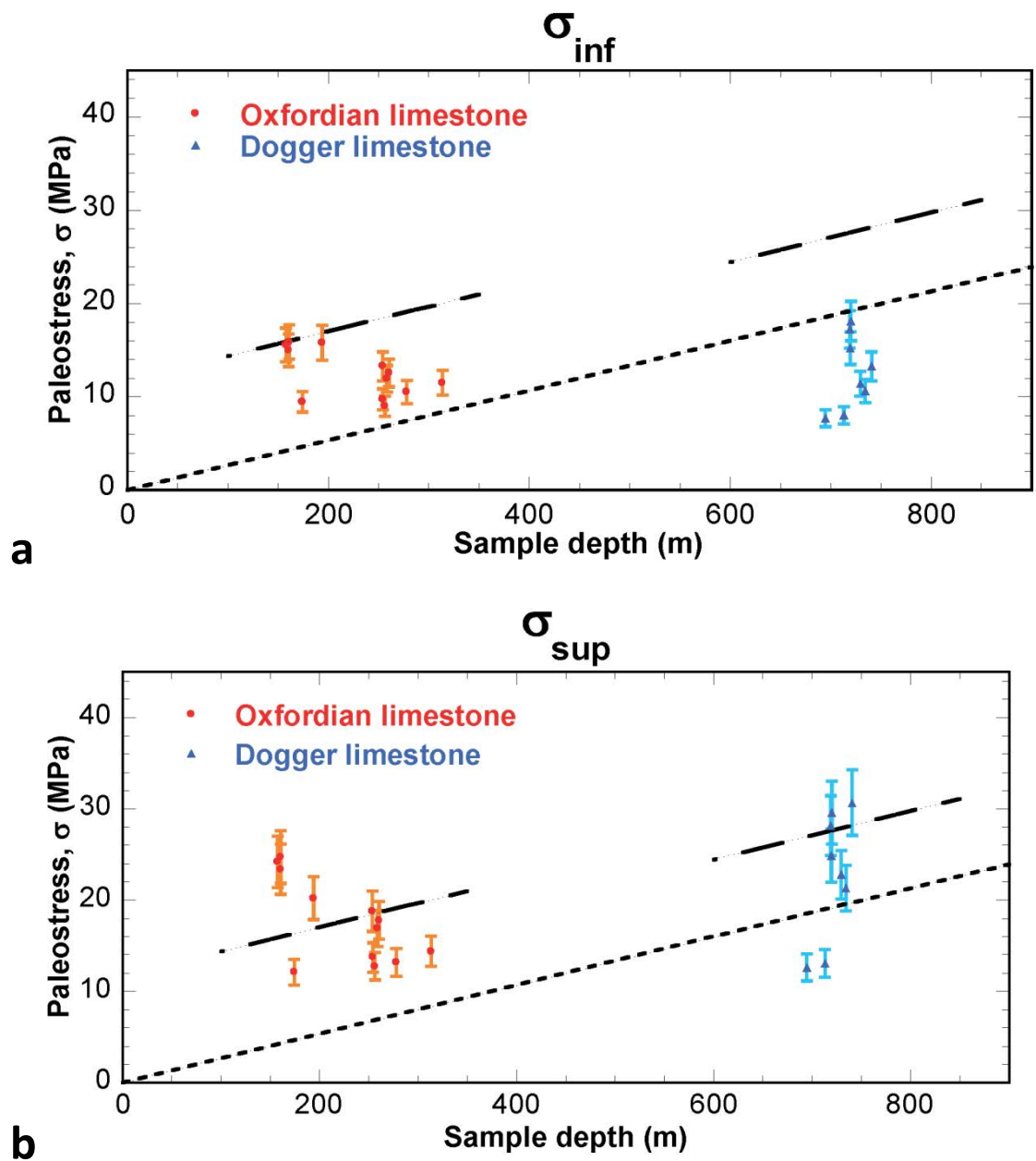

a

b

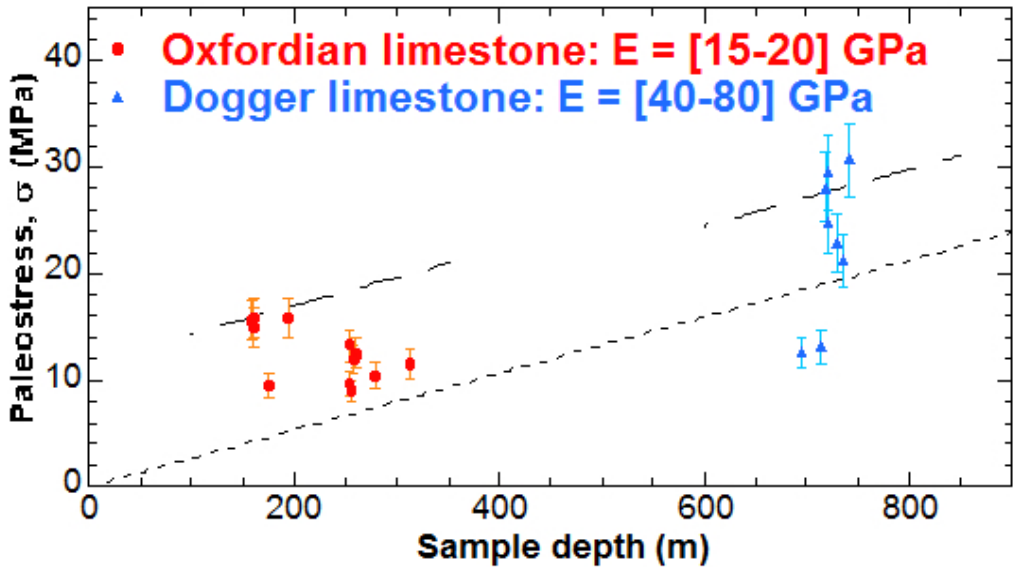

c

Figure 10: Results of the paleostress calculation for the Oxfordian and Dogger formations. The error bar on the data is 11.67%. a) The lower bound of the paleostress is calculated using Young's modulus equal to the lower limit for carbonates i.e. 15 GPa. b) The upper bound is calculated using the Young's modulus measured in the laboratory on selected samples. c) Associated Young's modulus for the most probable scenarios for the end of the stylolite evolution in the Oxfordian and Dogger horizons. For reference, we plotted on both graphs a line corresponding to the current lithostatic stress (i) and a line corresponding to the maximum overburden for both formations (ii - around 440 m of eroded ground for the borehole EST205 and 320 m for the borehole EST433) according to recent estimations [47].

Using the paleostress estimations of Fig. 10 we distinguished several groups at given stresses, corresponding to the groups observed previously in Fig. 9. Regarding the lower bound (Fig. 10a), the Oxfordian formation shows two groups: one at a stress of around 10 MPa, and one around 15 MPa. The group of stylolites at 15 MPa shows a stress equal to the one associated to the maximum overburden and thus it seems that the stylolites growth have been stopped when the overburden was the highest, thus the recorded stress is higher than the present value. The other group shows stresses between the current overburden stress and the maximum stress. This means that the stylolites stopped growing during either the burying phase or the erosion phase of the Paris Basin [48]. The Dogger formation shows three groups at 8, 11, and 17 MPa. All the corresponding stylolites indicate a stress lower than the prevailing stress, suggesting that, during their evolution, their formation was stopped before a stress equal to the present-day stress was reached. However, if we consider both groups close to the present vertical stress line, and if we convert the estimated stress at depth, a difficulty appears. Indeed, these rocks experienced a significant stress as they were buried to a depth of around 1000 m; therefore, it is not appropriate to consider a Young's modulus of 15 GPa for these rocks. The upper bound of the estimated stresses (Fig.10b) for the Oxfordian formation still shows two groups: one around 12 MPa, and another one at a stress higher than the maximum overburden associated stress. The latter case is not valid since it is impossible for the rock to have been buried below the maximum overburden depth. In that case, the Young's modulus may be too high and therefore a lower value has to be considered for these stylolites. For the group around 12 MPa, we are in the same scenario as before where the stylolite evolution is supposed to have ceased during the burying or erosion phase of the basin. However, in the Dogger formation, we observe two main groups compared to the lower bound of the estimated stress. A group of two stylolites indicates stresses of around 10 MPa, indicative of a halt in stylolite evolution at an early stage, at a stress inferior to the prevailing stress. The stylolites of the other group show a variation between the maximum overburden depth and the prevailing stress. In this case, the calculated stress is close to the present vertical stress. This suggests that (i) some stylolites are still active and, (ii) some stylolites progressively stopped growing at the beginning of the erosion phase.

To summarize, it seems that a low value of Young's modulus is more compatible with the Oxfordian formation, while the present Young's moduli are more appropriate for the Dogger formation (Fig. 10c). This leads to an average stresses of 10 and 15 MPa for the Oxfordian formation, and 10 and 30 to 22 MPa for the Dogger formation. In most cases, the stylolite evolution seems to have stopped at stresses between the maximum overburden stress and the present-day stress. This is corroborated by geochemical observations [30] showing two phases of stylolite reactivation in the late Cretaceous and the late Paleogene. These observations suggest that stylolite evolution was halted during the erosion phase after the late Cretaceous. With our method we have information about the time when stylolites stopped their growth. We can't infer their exact time of initiation. Our results combined with other studies on the site ([47]; [48]) allow us to suggest that stylolites probably stopped growing during the burial phase or the erosion phase. If we compare our results with the current stress field (Fig.1), we see that sedimentary stylolites can grow only in the Dogger formation below 500 m where $\sigma_V > \sigma_H$. This is in agreement with our interpretation where stylolites seem to have stop growing at depths between 400 and 800 m corresponding to stresses about 10 to 20 MPa (Fig. 10c) in the Oxfordian formation while we found some stylolites that may be still evolving in the Dogger.

5.2. *Tectonic stylolites*

As a pilot study we analysed tectonic stylolites within our borehole core samples. Locating tectonic stylolites was quite difficult as they are vertical or tilted, therefore the chance to have a non-destructive borehole crossing them are limited. Three vertical tectonic stylolites were found in the vicinity of the depths 175 m, 215 m, and 260 m for borehole EST205 and 179 m for borehole EST204 in the Oxfordian formation. In all cases, the vertical stylolites crossed the sedimentary stylolites that were distributed every 10 cm (Fig. 11a). Regarding the borehole EST205, the vertical stylolite at 260 m crossed the core at the edge and shows displacements at each sedimentary stylolite in its path. For this reason, we could not analyse a sufficiently long profile for that stylolite. The two other vertical stylolites crossed the core longitudinally and in the middle as for the tectonic stylolite in borehole EST204. We digitized and analysed the three of them. Only two stylolites showed a spectrum with the two-regime behaviour (Fig. 11b): the stylolite at 215 m (EST205) and the one at 179 m (EST204). The calculation of the associated horizontal paleostress (see details in Appendix 6) using the estimated cross-over lengths and with the low Young's modulus $E$ = 15 GPa and the measured Young's modulus $E$ = 31 GPa and $E$ = 25 GPa for boreholes EST205 and EST204, respectively, gives $10 \leq \sigma_H \leq 17$ MPa for both stylolites. We considered different depths, and thus different lithostatic stresses $\sigma_V$, depending on the tectonic events and burial history of the Paris Basin as shown in Brigaud et al. [48]. Considering the present stress field measurements (Fig.1), we see that the maximum horizontal principal stresses is $\sigma_H$ = 14.5 MPa which suggests the use of the upper bound of the Young's modulus leading to a stress $\sigma_H \approx 17$ MPa. Indeed, in the tectonic context in which those stylolites grew, the major horizontal principal stress may have been higher than the current stress. As a tectonic stylolite form only under a specific stress field (such as for mountain formation), and since previous studies showed numerous tectonic stages during Paleogene and Neogene periods (see for example André et al. [30]), these pilot results suggest that the tectonic stylolites initiated during one of these periods, and stopped when the major principal horizontal stress changed orientation. This pilot work resulted in a consistent estimation with respect to the present stress conditions (Fig. 1) and the history of the Paris Basin [48]. However, we were limited by the small number of tectonic stylolites found in the studied boreholes. More work is needed on a larger number of tectonic stylolites in the studied area. Our analysis suggests that the paleostress determination in various geological environments can be better determined when both horizontal and vertical non-destructive boreholes are available.

**Figure 11:**

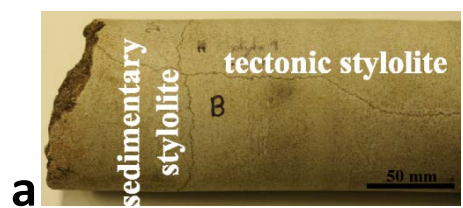

a

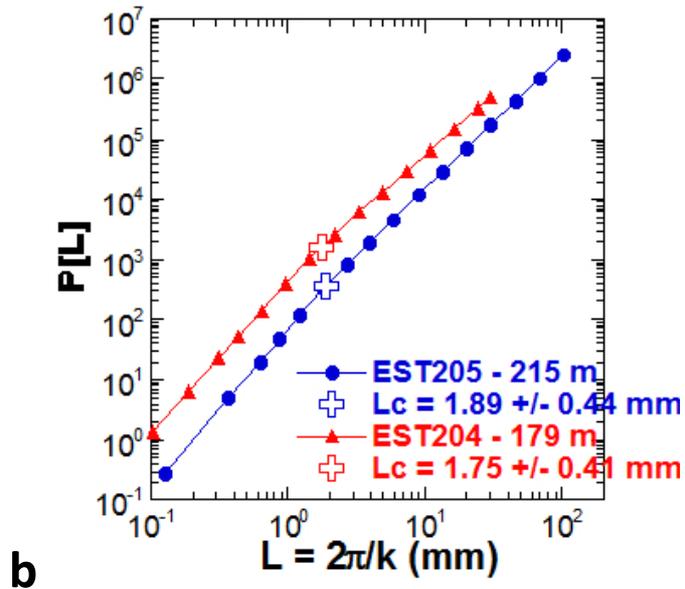

Figure 11: Tectonic stylolites analysis. a) Detail of one of the selected tectonic stylolite (core EST32219 in the vicinity of 215.2 m). The horizontal seam is the tectonic stylolite crossed by a sedimentary stylolite (vertical seam). b) Spectrum resulting from the analyses showing the two-regime behavior.

## 6. Conclusion

In this paper, we developed a rigorous methodology to infer paleostresses from cores taken from boreholes. Our procedure is based on a recent model [27] which relates the paleostresses to the morphology of stylolites. Using high-resolution photographs of stylolites, profiles were digitized and the resulting 1D morphologies were analysed by calculating the Fourier power spectrum method. The resulting spectra show a characteristic length, called the cross-over length, was used to estimate the associated paleostress. Numerous difficulties arose from the use of borehole core samples, such as geometrical constraints linked to their finite size. We show, on both natural and synthetic examples, that the morphology analysis performed on cylindrical contours and planar profiles yield comparable results. It is therefore possible to use the cylindrical contour to infer the paleostresses using a non-destructive procedure. We applied our new methodology to analyse a large number of stylolites in limestone formations surrounding the Andra URL at Bure (eastern Paris Basin, France). The paleostresses deduced from sedimentary stylolites are compatible with recent data on the evolution of the Paris basin:
- the stylolites from the Dogger horizon shows different growth history. One group is still active and recorded stresses from 20 to 24 MPa and two other groups are not active anymore and recorded paleostresses around 14 MPa and 28 MPa.
- the sedimentary stylolites from the Oxfordian horizon do not grow anymore and recorded paleostresses from 8 to 16 MPa.
- our results are in agreement with the present state of stress both in the Oxfordian and Dogger formations. While recent estimations showed that $\sigma_H > \sigma_V$ in the Oxfordian formation, enabling only the growth of tectonic stylolites above the Bure URL, the measurements suggested in contrast that $\sigma_V > \sigma_H$ for depths higher than 550 m, meaning that currently sedimentary stylolites can only grow in the Dogger horizon.

**Acknowledgments**
The authors want to thank Thierry Reuschlé, Mike Heap, Bertrand Renaudié, Jean-Daniel Bernard, Alain Steyer, Christian David, Philip Meredith, Marcus Ebner, Gianreto Manatschal and Yves Bernabé for technical support and fruitful discussions. They also thank two anonymous reviewers for constructive comments on the manuscript. This work is funded by Andra, with the support of FORPRO, NEEDS-MIPOR, the ITN FLOWTRANS and the CNRS INSU.

**Appendices**

*A1. Influence of the threshold used in the digitization procedure on the cross-over length*

To test the sensitivity to the thresholds in the digitization procedure we selected a segment of a stylolite from the Dogger formation (borehole EST433) from core EST44535 taken at a depth of 719.56 m. We used various thresholds on the grey levels (40, 60 and 80) and two different thresholds on the cluster size (800 and 1000). This stylolite was analysed with a threshold on grey levels of 60 and a threshold on the cluster size of binary objects of 1000 (arbitrary threshold depending on the continuity of the seam or on the size of heterogeneities present in the host rock) and the cross-over length was found to be 0.78 mm. In the following, we will study the different cases for the thresholds as [grey level threshold - cluster size threshold]. After the digitization we extracted the corresponding functions and we analysed them as described in section 4.2. Fig. A1a shows the results for the analyses with a threshold on the grey levels of 40. The seam of the stylolite is not well defined and thus the morphology is biased. We could not extract from the FPS the two-regimes behaviour. Fig. A1b and A1c show the results for the thresholds on the grey levels of 60 and 80. In the case of the threshold [60 - 800], we observe a two-regimes behaviour but the cross-over length is lower than what we expect and the Hurst exponent for large scales is a bit high. However, the thresholds [60 - 1000], [80 - 800] and [80 - 1000] show Hurst exponents and cross-over lengths close to what we expect. Therefore, if we underestimate the thresholds, the morphology is more affected than if we overestimate them. An overestimation of the thresholds induces more noise in the data but it is mixed up with the existing noise and it is therefore preferable than a lack in the data induced by an underestimation.

**Figure A1:**

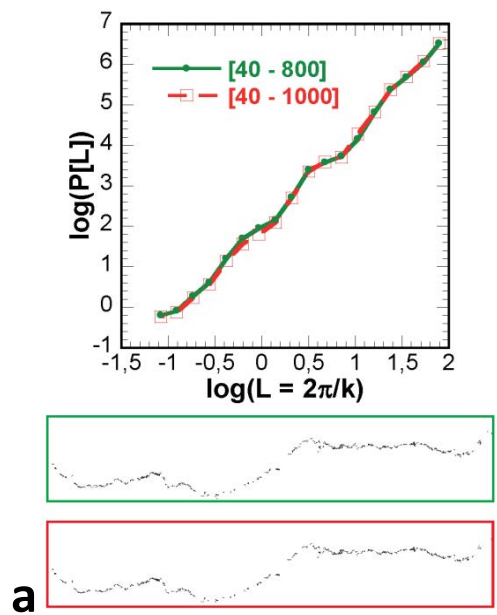

a

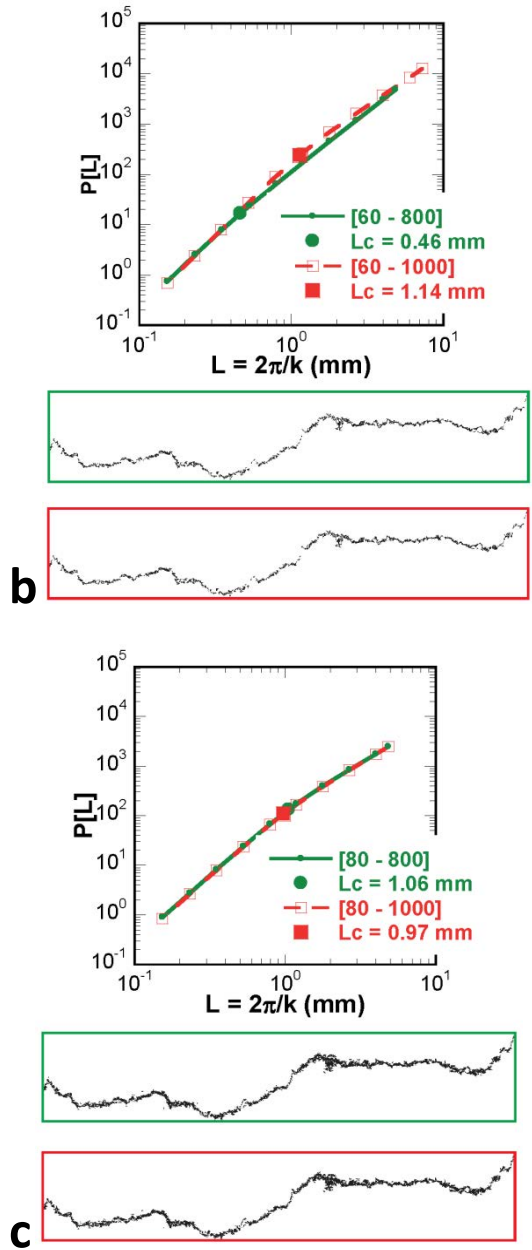

Figure A1: Sensitivity test of the procedure. Both value of thresholds are tested, the value on the grey levels and on the cluster size. a) Resampled spectrum data for threshold values of [40 - 800] and [40 - 1000]. The stylolite seams are framed with the corresponding colour. b) Modeled data and cross-over lengths for threshold values of [60 - 800] and [60 - 1000]. The stylolite seams are framed with the corresponding colour. c) Modeled data and cross-over lengths for threshold values of [80 - 800] and [80 - 1000]. The stylolite seams are framed with the corresponding colour.

*A2. Linear interpolation*

As explained in step v of the digitizing procedure (section 4), we interpolate the discontinuities for the binarized profiles using a linear interpolation. The lines are plotted between the middle of the edge of the involved clusters. We compared the interpolation percentage and the spectrum resulting from the morphology analysis for each stylolite to assess if the ill-defined spectrums and the interpolation percentage are linked. Figure A2 shows the interpolation percentage as a function of the cross-over length for the Oxfordian (Fig. A2a) and Dogger stylolites (Fig. A2b). The stylolites where no cross-over lengths were found are symbolised by a 10 mm cross-over length. For both horizons, there is no obvious relation between the interpolation percentage and the cross-over length and thus the spectrum.

We can conclude that the use of this linear interpolation does not induce any discrepancy in the extracted morphology.
**Figure A2:**

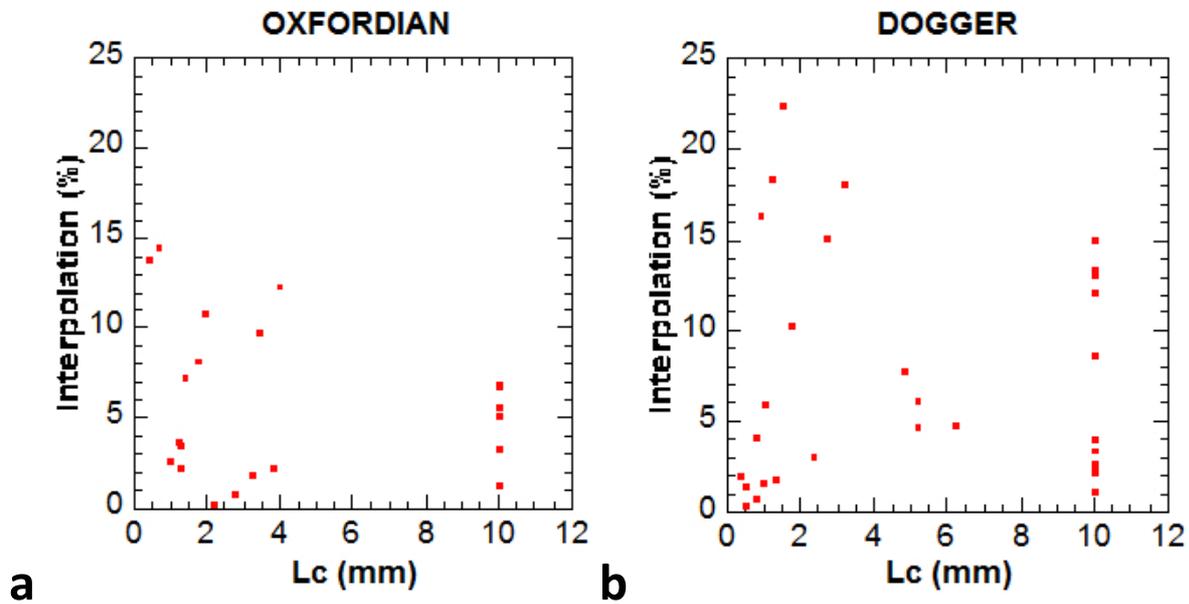

Figure A2: Interpolation percentage as a function of the cross-over lengths obtained by the morphology analyses of (a) Oxfordian stylolites and (b) Dogger stylolites. The 10 mm cross-over length corresponds to stylolites which analysis resulted in ill-defined or one scaling law spectrum.

*A3. Repeatability of the morphological analysis*
We tested the repeatability of the method by performing the analysis several times on the same stylolite. We repeated the whole procedure from the digitization to the FFT analysis. Regarding the digitization, Fig. A3a shows the extracted function from the first analysis in blue and from the second analysis several months later in red. We see that they are almost identical. To assess the differences between both traces we calculated the absolute difference between both traces (Fig. A3b). The maximum difference corresponds to 18.3 % of the total amplitude of the stylolite while the mean of the difference corresponds to less than 1 % of the amplitude. Regarding the analyses results, Fig. A3c shows the spectrum of the first analysis and Fig. A2d shows the spectrum obtained after the second analysis. We see that the Hurst exponents are almost the same and that the cross-over lengths have a difference of 24.8 % which is of the order of the uncertainty of the method.
**Figure A3:**

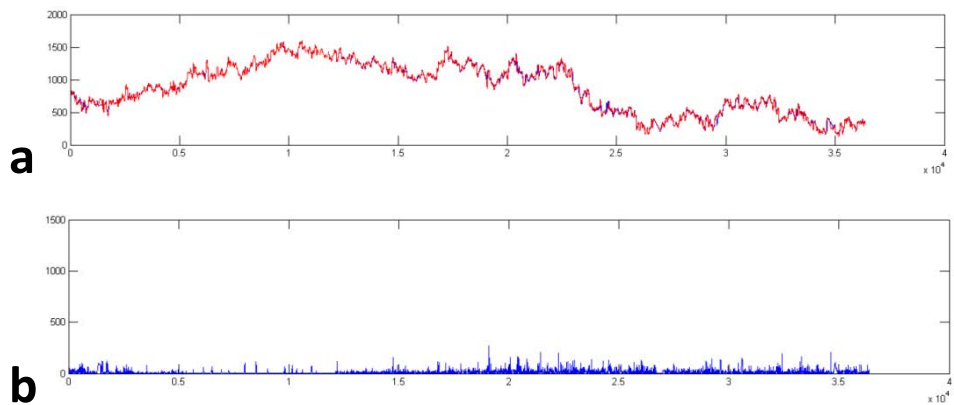

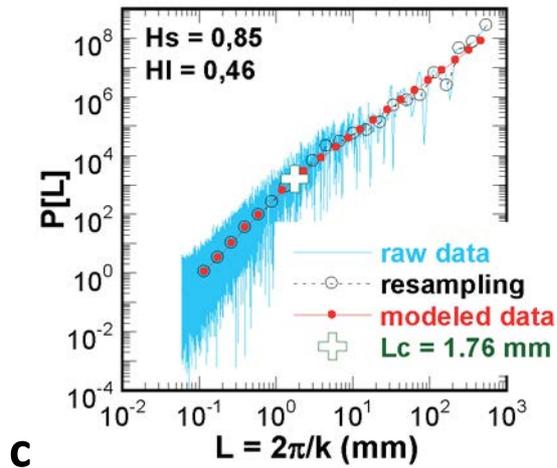

c

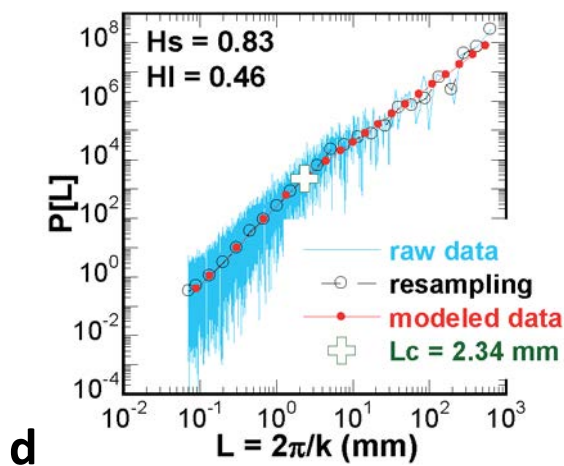

d

Figure A3: Repeatability analysis of the method performed by analysing twice the same stylolite. a) Profiles of the first (blue) and the second (red) analysis performed independently. They are almost identical. b) Differences between both traces. The mean difference represents less than 1 % of the total amplitude of the stylolite. Spectrums resulting from c) the first analysis and d) the second analysis. The Hurst exponents are almost the same while the cross-over lengths difference is in the error bar of the method.

*A4. Morphological comparison between a circular profile and a planar profile*
In addition to the analysis shown on a natural stylolite from the Dogger formation (section 4.3), we created a synthetic stylolitic surface ([43], [44]), with known Hurst exponents (1 for the small-scale and 0.5 for the large-scale) and a cross-over length of 1 mm. Our analysis consisted of generating a white noise (here with values between -1 and 1 and centred in 0) on which we applied a fast Fourier transform – FFT. We imposed the Hurst exponents for the small- and large-scale in the frequency domain by a discrimination over *k*, the wave-number, such as for $k > k_c = 2\pi/L_c, H_s = 1$ for the small-scale and vice versa for large-scale with $H_l = 0.5$. We then came back to the spatial domain by performing a reversed FFT, and we obtained the synthetic stylolitic surface (Fig. A4a). The results (Fig. A4b) show that using the external profile, or the planar profile, gives non-distinguishable results, well within the error bars (they are closer to each other than the error bars of 23.34 % that correspond to the dispersion of the cross-over between independent planar profiles analysed in the same surface). Thus we characterise the morphology of stylolites using the external profiles of borehole core samples in a non-destructive manner.
**Figure A4:**

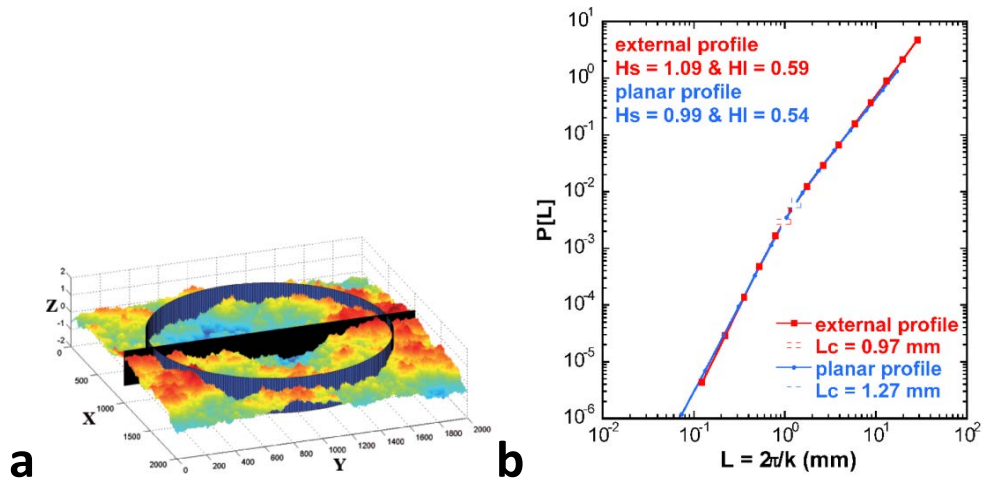

Figure A4: Test analyses on a synthetic stylolitic surface. a) Synthetic stylolitic surface. The black surface cutting along the diameter simulates a planar profile. The blue surface cutting along a perimeter simulates an external contour of a core. b) Results of the test analyses for the planar profile and the external contour. The Fourier power spectrum is represented as a function of the wave-length.

*A5. Assessment of anisotropy for the sedimentary stylolites*

To check the existence of anisotropy for the sedimentary stylolites profiles and validate the use of one cross-over length in our interpretations (section 5), we analysed the stylolite morphology over different azimuths. We divided the stylolite analysed in Fig. 6 in four parts that we named part 1/4, part 2/4, part 3/4 and part 4/4. The different parts were analysed with the procedure described in section 4. Fig. A5 shows the result of these analyses and the four analysed parts. Considering the error bars, no significant differences was observed between the different cross-over lengths for the different parts (< 25%). These results confirm the overall isotropy of our problem, in contrast with previous results of Ebner et al. (2010) who found an anisotropy of the order of 30 to 60 % for tectonic stylolites.

**Figure A5:**

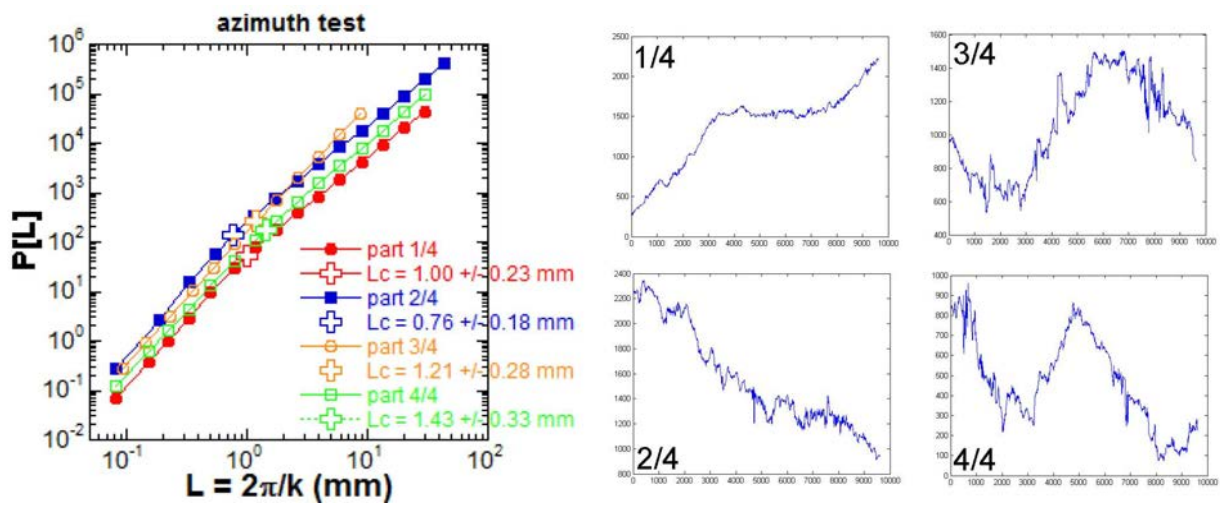

Figure A5: Assessment of anisotropy for sedimentary stylolites. A stylolite from the Dogger formation (see Figure 5) was split in 4 parts. The morphology of each part was analysed. No significant anisotropy for the estimated cross-over lengths was observed.

*A6. Paleostress calculation for tectonic stylolites*

From Equation (1), the following relationship is extracted: $L_c = \gamma * E / (\beta * P * \sigma_s)$ where $L_c$ is the cross-over length, $\gamma$ is the surface tension, $\beta = \frac{[2\nu(1-2\nu)]}{\pi}$ is a dimensionless parameter, $E$ and $\nu$ are the elastic parameters (Young's modulus and Poisson's ratio, respectively) and $P$ and $\sigma_S$ are the mean and differential stress, respectively. As we have no clue about the stress field during the growth of the tectonic stylolites, we made the assumption that stresses in the mean plane of the stylolites are isotropic: $\sigma_{zz} = \sigma_{xx}$ with $\sigma_{yy}$ the major principal stress. Using Equation (1), the mean stress $P = (\sigma_{yy} + 2\sigma_{zz})/3$ and the differential stress $\sigma_S = \sigma_{yy} - \sigma_{zz}$, we obtain the following quadratic equation: $\sigma_{yy}^2 + \sigma_{zz} * \sigma_{yy} - 2\sigma_{zz}^2 = 3\gamma E / \beta L_c$ where $\sigma_{yy}$ is the unknown parameter. We solved the equation and determined the relationship: $\sigma_{yy} = \left(\sqrt{9\sigma_{zz}^2 + \left(12\gamma E / \beta L_c\right)} - \sigma_{zz}\right)/2$. From this relationship and knowing the burial history of the basin, which allows to estimate intervals for $\sigma_{zz}$, we assessed intervals for the major principal stress corresponding to the end of the tectonic stylolite growth.